\documentclass[a4paper,11pt]{article}
\usepackage{pos}
\usepackage{caption}
\usepackage{subcaption}

\title{W-boson angular coefficients at LHC at high precision}

\author*[a]{Timea Vitos}

\affiliation[a]{Theoretical particle physics, Lund University, \\ Sölvegatan 14A, SE-223 62, Lund, Sweden}

\emailAdd{timea.vitos@thep.lu.se}

\abstract{We present state-of-the-art high-precision theory predictions for the dominant angular coefficients parametrizing the spin-correlations in the production and decay of a W-boson produced at transverse momentum larger than 30 GeV. The computation at NNLO QCD and NLO EW accuracy are combined to obtain differential distributions in the W-boson transverse momentum and rapidity. The found results show up to 10~\% corrections in certain regions of phase space, while the scale bands are significantly reduced as compared to NLO QCD.}

\FullConference{%
  The Tenth Annual Conference on Large Hadron Collider Physics - LHCP2022\\
  16-20 May 2022\\
  online
}

\begin{document}
\maketitle

\section{Introduction}

The W-boson mass is one of the fundamental parameters of the Standard Model, which, through its connection to the other electroweak parameters and thus to the electroweak symmetry breaking, is an important key to detecting beyond Standard Model physics. One such process from which the mass can be measured is the W-boson production at non-zero transverse momentum (W+jet process). The mass parameter is however not measured directly, due to the inherent difficulty of measuring the neutrino, but rather through template fits from theory predictions. 

In order to reduce both the theoretical and experimental systematic uncertainties, prior to the W-boson mass template fits, a theory-data comparison is made to the similar Z+jet process \cite{ATLAS:2017rzl}. Eventual discrepancies between theory and data in the Z+jet process is then translated to the W+jet process. In order to do this efficiently, and to reduce statistical fluctuations in the theory computations, the five-dimensional phase space of the V+jet process can be translated into a three-dimensional one, together with eight angular coefficient functions, describing the decay of the vector boson. Despite the  similarities, the two processes are still not the same: different parton luminosities contribute, the scales are different, and the EW corrections differ.  As illustration, we show the comparison of the A$_4$ coefficient, differentially in the vector boson p$_T$ at NLO QCD and at NLO EW\footnote{For the EW K-factor, the LO results are computed with a one-loop $\rho$-parameter for a fair comparison to the NLO EW results.} precision in Fig. \ref{fig:diag}. It is thus necessary to predict both of the processes and the corresponding angular coefficients with the current state-of-the-art accuracy. The Z+jet process has been presented at NNLO QCD \cite{Gauld:2017tww} and NLO EW \cite{Frederix:2020nyw}. In the current work, we present a combination of NNLO QCD and NLO EW corrections to the W-boson related process \cite{Pellen:2022fom} for the 13 TeV LHC. 

\section{Theory and computational setup}

The differential cross section in terms of the eight angular coefficients is

\begin{eqnarray}
\begin{split}
\label{eq:coef}
\frac{\text{d}\sigma}{\text{d}p_{{\rm T},W}\,\text{d}y_W\,\text{d}m_{\ell\nu}\,\text{d}\Omega}=
    &\frac{3}{16\pi}\frac{\text{d}\sigma^{U+L}}{\text{d}p_{{\rm T},W}\,\text{d}y_W\,\text{d}m_{\ell\nu}} \bigg((1+\cos^2\theta)+{\rm A}_0\frac{1}{2}(1-3\cos^2 \theta) \\
& +{\rm A}_1 \sin 2\theta \cos \phi + {\rm A}_2 \frac{1}{2}\sin^2\theta \cos 2\phi +{\rm A}_3\sin \theta \cos \phi +{\rm A}_4 \cos \theta \\
 & +{\rm A}_5 \sin^2 \theta \sin2\phi+{\rm A}_6 \sin2\theta \sin\phi +{\rm A}_7 \sin \theta \sin \phi\bigg) ,
 \end{split}
\end{eqnarray}

in which the (charged) lepton angles $\theta,\phi$ are suitably evaluated in the Collins-Soper frame \cite{Collins:1977iv}, and $\sigma^{U+L}$ refers to the unpolarized normalizing cross section. 

\begin{figure}
\centering
        \begin{subfigure}{0.49\textwidth}
        \center
                 \includegraphics[width=0.5\textwidth]{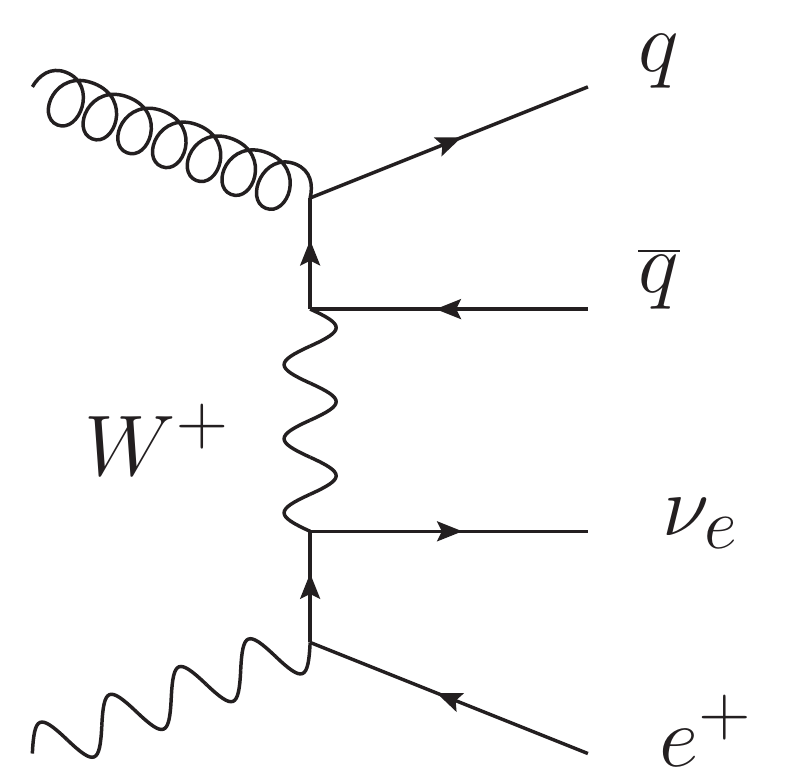}
        \end{subfigure}
        \hfill
        \begin{subfigure}{0.49\textwidth}
                 \includegraphics[width=0.85\textwidth]{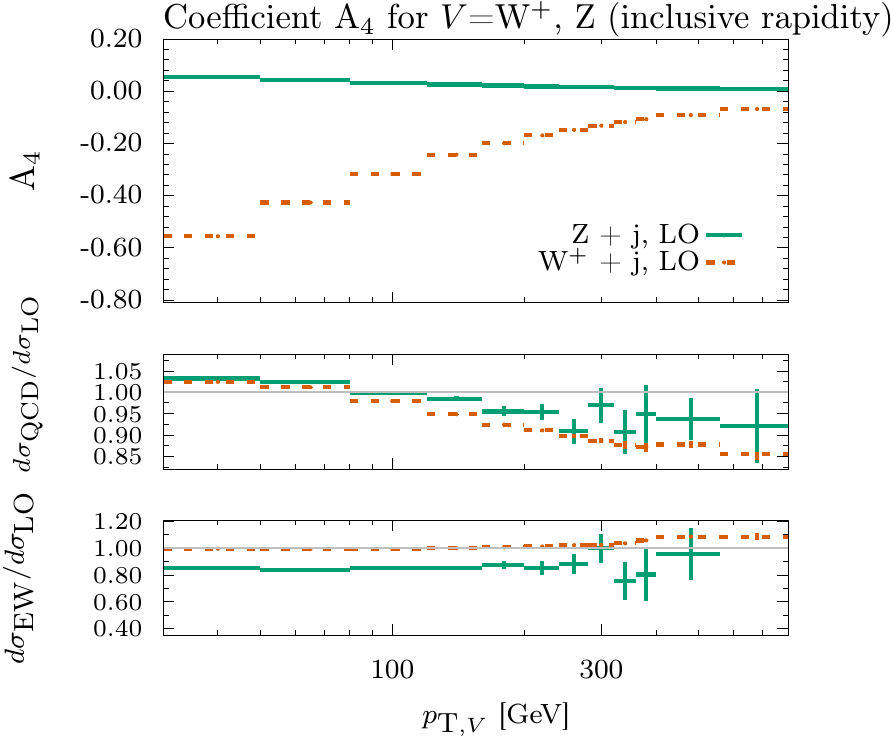}
        \end{subfigure}
        \caption{\textbf{Left:} Exemplary Feynman diagram of photon-induced contributions featuring double-soft singularities at the current perturbative order. \textbf{Right:} The A$_4$ coefficient for the Z+jet process and W$^+$ +jet (upper panel), the corresponding NLO QCD K-factors (middle panel) and the NLO EW K-factors (lower panel).}
        \label{fig:diag}
\end{figure}

This work considers a combination of higher-order corrections, separately at NNLO QCD ($\mathcal{O}(\alpha_s^3 \alpha^2)$), computed using the {\sc Stripper} framework~\cite{Czakon:2010td,Czakon:2011ve,Czakon:2014oma,Czakon:2019tmo} within the narrow-width approximation\footnote{The off-shell effects at NLO QCD have been cross-checked to be negligible.} and  NLO EW ($\mathcal{O}(\alpha_s\alpha^3)$), using \textsc{MadGraph5\_aMC@NLO}~\cite{Alwall:2014hca,Frederix:2018nkq}. The mixed QCD+EW correction at the combined order however is ambiguously defined, as the angular coefficients are defined as ratios of observables,
\begin{eqnarray}
\text{A}_j^{\rm def} = \frac{\sigma_{\rm num}}{\sigma_{\rm den}}.
\end{eqnarray}
In this work, we compare both an unexpanded ratio (denoted as \emph{def}) and an expanded ratio in the strong coupling constant (denoted as \emph{exp}),
\begin{eqnarray}
{\rm A}^{\rm exp}_j = A+\alpha_s B + \alpha_s^2 C.
\end{eqnarray}
 The electroweak corrections are then included as $K$-factors at the level of the coefficients, 
 \begin{eqnarray}
{\rm A}_{j, {\rm QCD + EW}}^{\rm def/exp}  = K_\text{EW} \times {\rm A}_{j}^{\rm def/exp} 
\end{eqnarray}
 in order to remedy a soft divergence present in the calculations at this order\footnote{The double-soft IR singularity is not canceled at the separate $\mathcal{O}(\alpha_s\alpha^3)$ order, as this requires mixed QCD+EW loop diagrams to be included.}, portrayed in the left of Fig. \ref{fig:diag}.

In the electroweak sector, the complex-$G_{\mu}$ and the complex-mass-scheme used, and final-state photons are recombined with the charged leptons. As the central scales, we choose the transverse energy of the W-boson and as customary perform a 7-point variation on the numerator and denominator, and a 31-point variation on the ratio.

\section{Results}

\begin{figure}
\centering
        \begin{subfigure}{0.49\textwidth}
        \center
                 \includegraphics[width=0.68\textwidth]{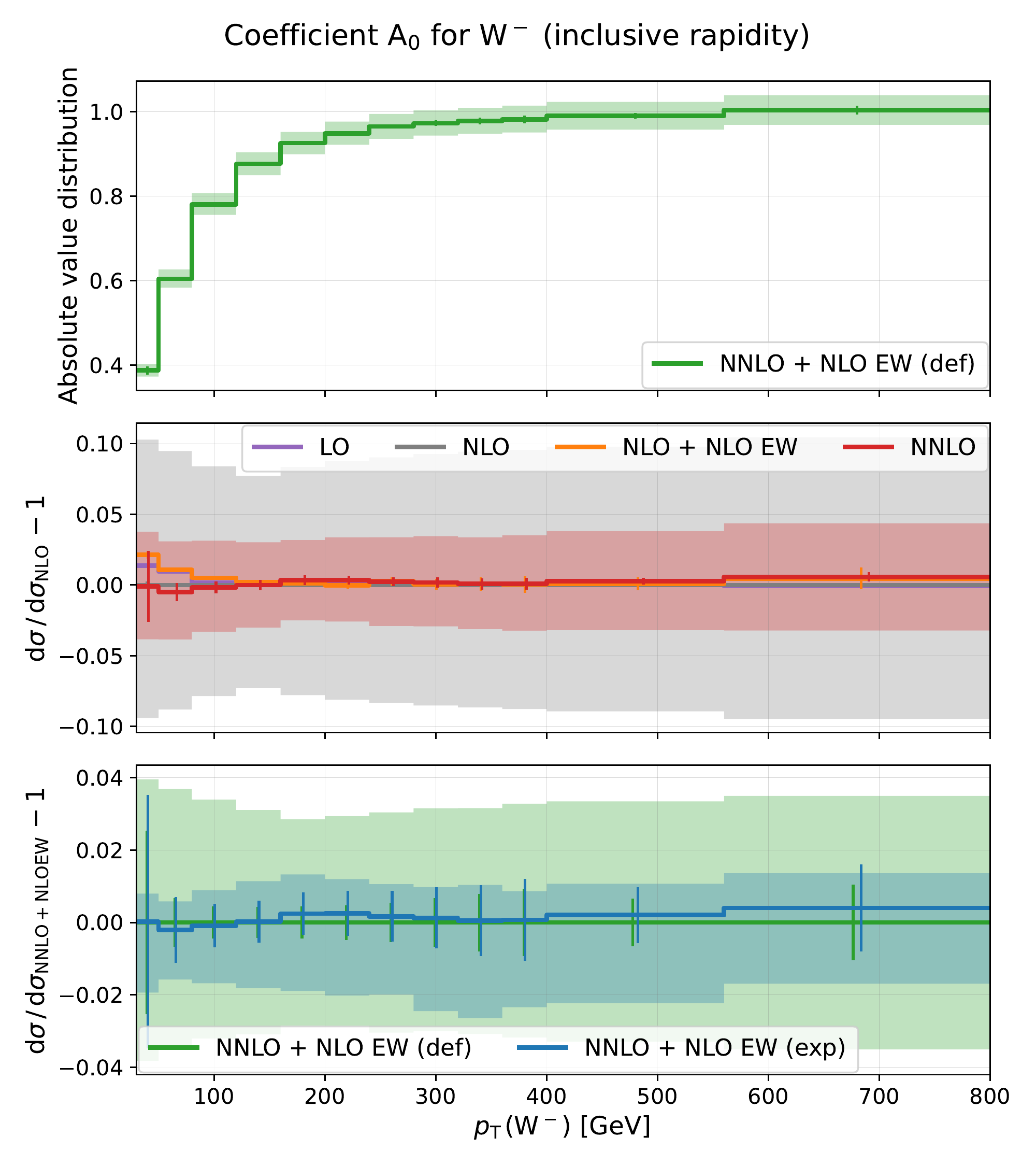}
        \end{subfigure}
        \hfill
        \begin{subfigure}{0.49\textwidth}
        \center
                 \includegraphics[width=0.68\textwidth]{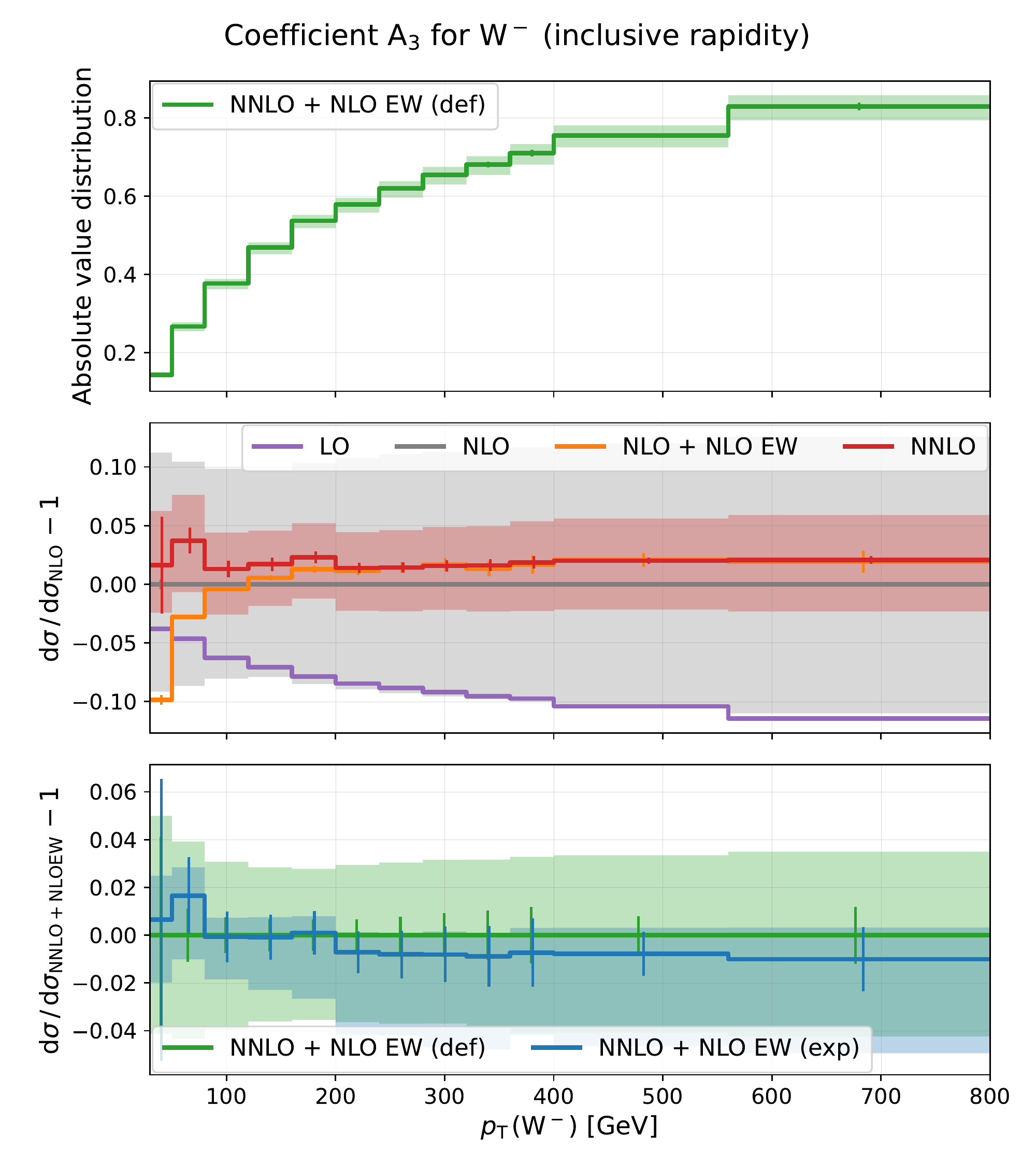}
        \end{subfigure}
\caption{The A$_0$ (\textbf{left}) and A$_3$ \textbf{(right)} coefficients, inclusive in the W-boson rapidity.}
        \label{fig:A0_A3_yinc}
\end{figure}

We show the A$_0$ and A$_3$ coefficients, inclusive in the W-boson rapidity, in Fig. \ref{fig:A0_A3_yinc} for W$^-$ production\footnote{The corresponding results for the plus signature behave similarly for all coefficients except the parity-odd A$_3$ and A$_4$, in which cases the distributions change sign.}. The factor $\sim$2 decrease in the scale uncertainty at NNLO QCD as compared to the NLO QCD result is visible for both of these (and  all the other) coefficients. In the lower panels, we make a comparison between the expanded and unexpanded versions of the coefficients, where we note that while the central values show negligible difference, the scale bands are a factor $\sim$2  smaller for the expanded versions.

\begin{figure}
\centering
        \begin{subfigure}{0.49\textwidth}
        \center
                 \includegraphics[width=0.68\textwidth]{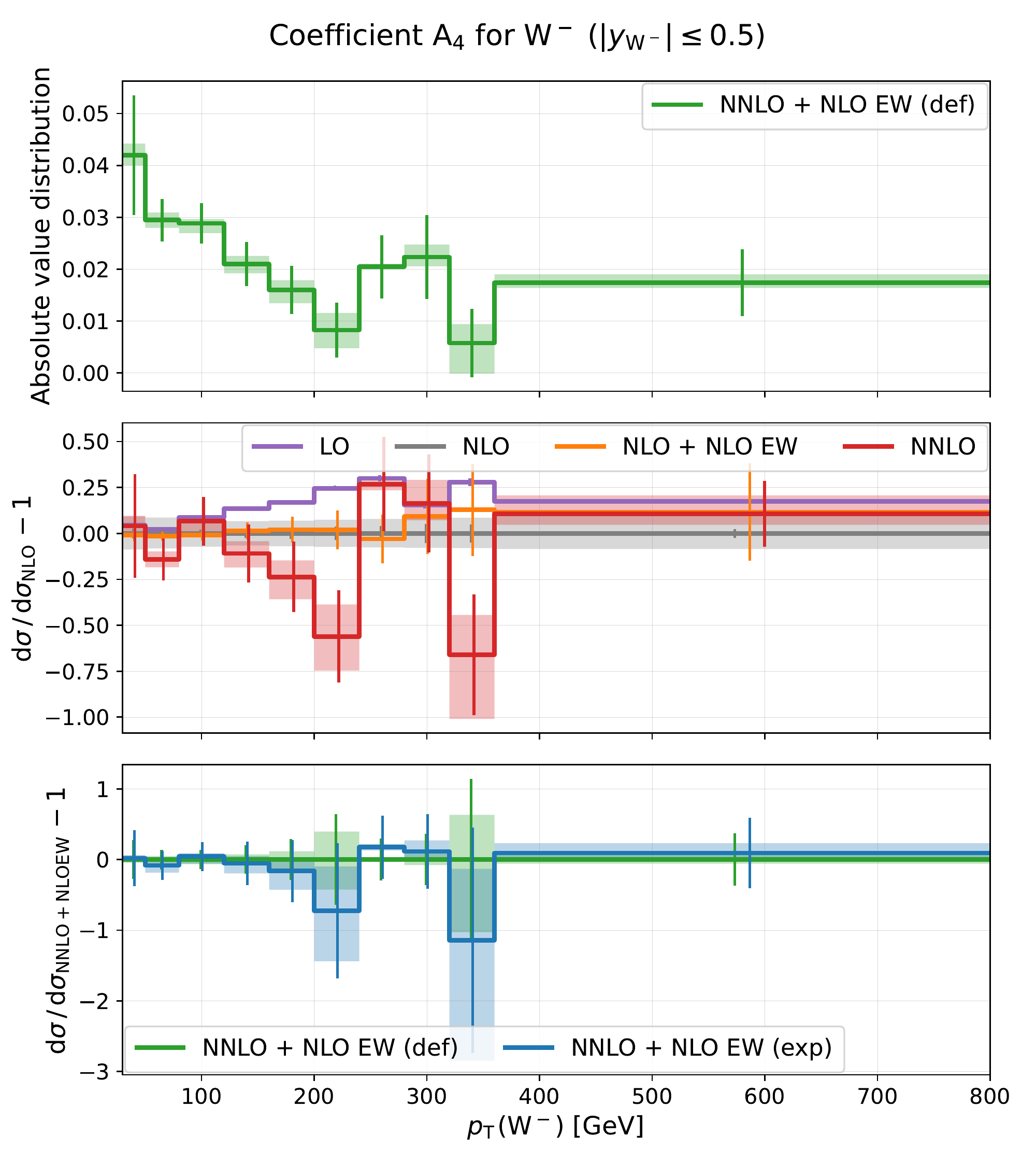}
        \end{subfigure}
        \hfill
        \begin{subfigure}{0.49\textwidth}
        \center
                 \includegraphics[width=0.68\textwidth]{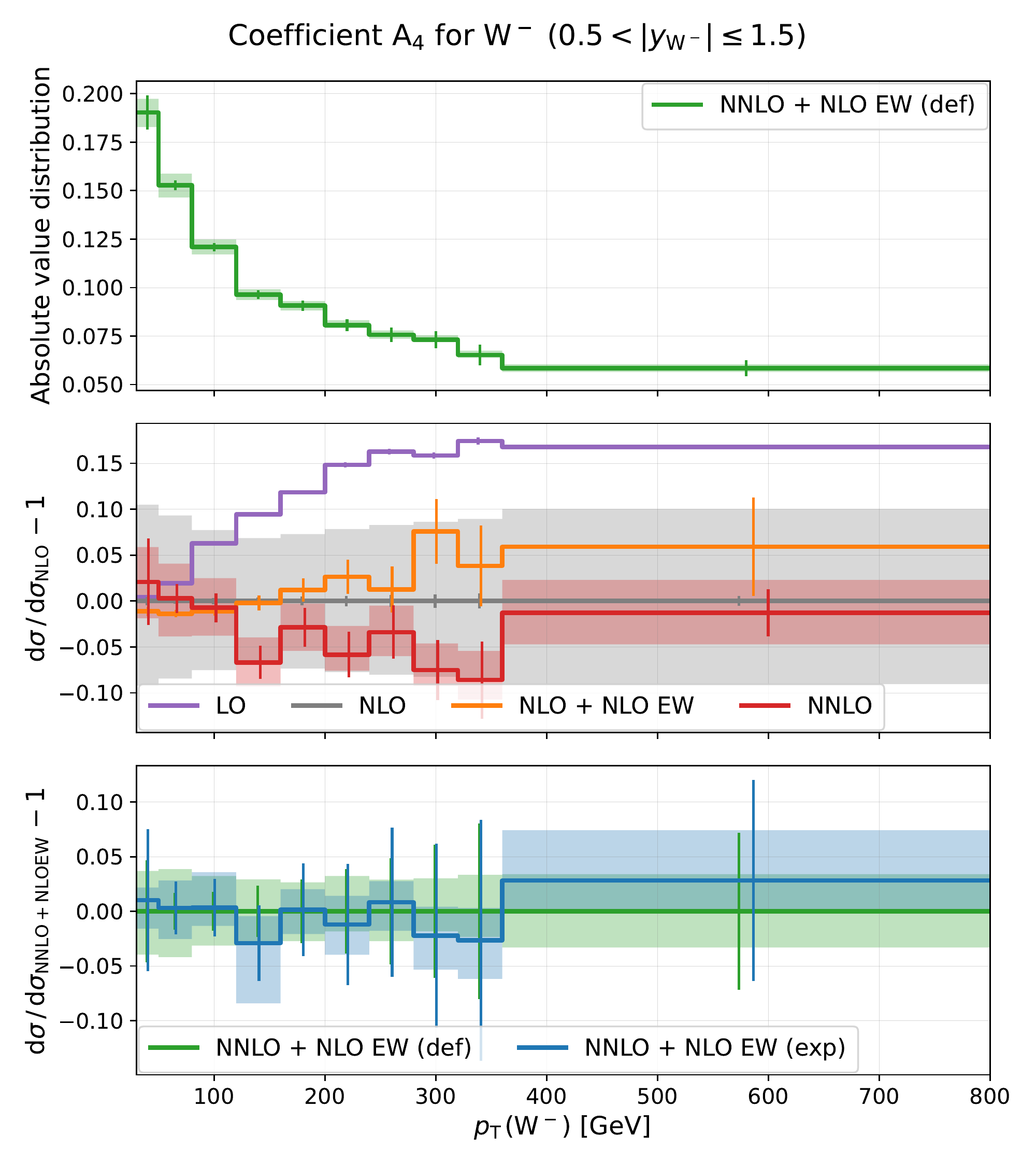}
        \end{subfigure}
\caption{The A$_4$ coefficient in the central-rapidity region \textbf{(left)} and in the mid-rapidity region \textbf{(right)}. }
                \label{fig:y1}
\end{figure}

Regarding the rapidity-dependence of the coefficients, in Fig. \ref{fig:y1} are shown the A$_4$ coefficient in two rapidity bins. Similarly, A$_1$ and A$_3$ are found to have significant rapidity-dependence, while the A$_0$ and A$_2$ coefficients remain rapidity-independent.

The decomposition into angular coefficients as in Eq. \eqref{eq:coef} is strictly only valid when the boson decay is a 2-body decay, which is broken when electroweak corrections are considered. This effect was examined in the work by performing a reweighted event generation for LO events, reweighted with NLO EW accurate coefficients. We found this non-completeness electroweak effect to be negligible. 

\section{Conclusion}

We found moderate corrections at NNLO QCD and NLO EW for the angular coefficients, while the scale uncertainties were found to significantly decrease. This motivates well these high-precision predictions to be the starting point for an experimental determination of these coefficients, and alongside it, a more accurate W-boson mass measurement.
\acknowledgments

This work was done together with Mathieu Pellen, Rene Poncelet and Andrei Popescu. Thanks to Rikkert Frederix for his valuable input.

\newpage
\bibliographystyle{JHEP}
\bibliography{PoS_Vitos_LHCP2022}

\providecommand{\href}[2]{#2}\begingroup\raggedright\begin{thebibliography}{10}

\bibitem{ATLAS:2017rzl}
{\scshape ATLAS} collaboration, \emph{{Measurement of the $W$-boson mass in pp
  collisions at $\sqrt{s}=7$ TeV with the ATLAS detector}},
  \href{https://doi.org/10.1140/epjc/s10052-017-5475-4}{\emph{Eur. Phys. J. C}
  {\bfseries 78} (2018) 110}
  [\href{https://arxiv.org/abs/1701.07240}{{\ttfamily 1701.07240}}].

\bibitem{Gauld:2017tww}
R.~Gauld, A.~Gehrmann-De~Ridder, T.~Gehrmann, E.W.N.~Glover and A.~Huss,
  \emph{{Precise predictions for the angular coefficients in Z-boson production
  at the LHC}}, \href{https://doi.org/10.1007/JHEP11(2017)003}{\emph{JHEP}
  {\bfseries 11} (2017) 003}
  [\href{https://arxiv.org/abs/1708.00008}{{\ttfamily 1708.00008}}].

\bibitem{Frederix:2020nyw}
R.~Frederix and T.~Vitos, \emph{{Electroweak corrections to the angular
  coefficients in finite-$p_T$Z-boson production and dilepton decay}},
  \href{https://doi.org/10.1140/epjc/s10052-020-08513-7}{\emph{Eur. Phys. J. C}
  {\bfseries 80} (2020) 939}
  [\href{https://arxiv.org/abs/2007.08867}{{\ttfamily 2007.08867}}].

\bibitem{Pellen:2022fom}
M.~Pellen, R.~Poncelet, A.~Popescu and T.~Vitos, \emph{{Angular coefficients in
  $\hbox {W}+\hbox {j}$ production at the LHC with high precision}},
  \href{https://doi.org/10.1140/epjc/s10052-022-10641-1}{\emph{Eur. Phys. J. C}
  {\bfseries 82} (2022) 693}
  [\href{https://arxiv.org/abs/2204.12394}{{\ttfamily 2204.12394}}].

\bibitem{Collins:1977iv}
J.C.~Collins and D.E.~Soper, \emph{{Angular Distribution of Dileptons in
  High-Energy Hadron Collisions}},
  \href{https://doi.org/10.1103/PhysRevD.16.2219}{\emph{Phys. Rev. D}
  {\bfseries 16} (1977) 2219}.

\bibitem{Czakon:2010td}
M.~Czakon, \emph{{A novel subtraction scheme for double-real radiation at
  NNLO}}, \href{https://doi.org/10.1016/j.physletb.2010.08.036}{\emph{Phys.
  Lett. B} {\bfseries 693} (2010) 259}
  [\href{https://arxiv.org/abs/1005.0274}{{\ttfamily 1005.0274}}].

\bibitem{Czakon:2011ve}
M.~Czakon, \emph{{Double-real radiation in hadronic top quark pair production
  as a proof of a certain concept}},
  \href{https://doi.org/10.1016/j.nuclphysb.2011.03.020}{\emph{Nucl. Phys. B}
  {\bfseries 849} (2011) 250}
  [\href{https://arxiv.org/abs/1101.0642}{{\ttfamily 1101.0642}}].

\bibitem{Czakon:2014oma}
M.~Czakon and D.~Heymes, \emph{{Four-dimensional formulation of the
  sector-improved residue subtraction scheme}},
  \href{https://doi.org/10.1016/j.nuclphysb.2014.11.006}{\emph{Nucl. Phys. B}
  {\bfseries 890} (2014) 152}
  [\href{https://arxiv.org/abs/1408.2500}{{\ttfamily 1408.2500}}].

\bibitem{Czakon:2019tmo}
M.~Czakon, A.~van Hameren, A.~Mitov and R.~Poncelet, \emph{{Single-jet
  inclusive rates with exact color at $ \mathcal{O} $ ($ {\alpha}_s^4 $)}},
  \href{https://doi.org/10.1007/JHEP10(2019)262}{\emph{JHEP} {\bfseries 10}
  (2019) 262} [\href{https://arxiv.org/abs/1907.12911}{{\ttfamily
  1907.12911}}].

\bibitem{Alwall:2014hca}
J.~Alwall, R.~Frederix, S.~Frixione, V.~Hirschi, F.~Maltoni, O.~Mattelaer
  et~al., \emph{{The automated computation of tree-level and next-to-leading
  order differential cross sections, and their matching to parton shower
  simulations}}, \href{https://doi.org/10.1007/JHEP07(2014)079}{\emph{JHEP}
  {\bfseries 07} (2014) 079} [\href{https://arxiv.org/abs/1405.0301}{{\ttfamily
  1405.0301}}].

\bibitem{Frederix:2018nkq}
R.~Frederix, S.~Frixione, V.~Hirschi, D.~Pagani, H.S.~Shao and M.~Zaro,
  \emph{{The automation of next-to-leading order electroweak calculations}},
  \href{https://doi.org/10.1007/JHEP11(2021)085}{\emph{JHEP} {\bfseries 07}
  (2018) 185} [\href{https://arxiv.org/abs/1804.10017}{{\ttfamily
  1804.10017}}].

\end{thebibliography}\endgroup

\end{document}